\documentclass[aps,preprint]{revtex4-1}%
\usepackage{amsfonts}
\usepackage{amsmath}
\usepackage{amssymb}
\usepackage{graphicx}
\usepackage{hyperref}
\usepackage{natbib}
\usepackage{color}
\providecommand{\U}[1]{\protect\rule{.1in}{.1in}}

\begin{document}
\preprint{HEP/123-qed}
\title[Short title for running header]{About Superrotation in Venus}
\author{D.J. Cirilo-Lombardo}
\affiliation{CONICET- Universidad de Buenos Aires. Instituto de F\'\i sica del Plasma (INFIP). Buenos Aires, Argentina} 
\affiliation{Universidad de Buenos Aires. Facultad de Ciencias Exactas y Naturales. Departamento de F\'\i sica. Buenos Aires, Argentina}
\affiliation{Bogoliubov Laboratory of Theoretical Physics, Joint Institute for Nuclear Research, 141980 Dubna, Russian Federation}
\author{M. Mayochi}
\affiliation{Universidad de Buenos Aires. Facultad de Ciencias Exactas y Naturales. Departamento de F\'\i sica. Buenos Aires, Argentina}
\author{F.O. Minotti}
\affiliation{Universidad de Buenos Aires. Facultad de Ciencias Exactas y Naturales. Departamento de F\'\i sica. Buenos Aires, Argentina}
\affiliation{CONICET- Universidad de Buenos Aires. Instituto de F\'\i sica del Plasma (INFIP). Buenos Aires, Argentina}
\pacs{PACS number}
\author{C.D.Vigh}
\affiliation{Instituto de Ciencias, Universidad Nacional de General Sarmiento, 
Los Polvorines, Argentina}
\affiliation{Universidad de Buenos Aires. Facultad de Ciencias Exactas y Naturales. Departamento de F\'\i sica. Buenos Aires, Argentina}
\affiliation{CONICET- Universidad de Buenos Aires. Instituto de F\'\i sica del Plasma (INFIP). Buenos Aires, Argentina}

\begin{abstract}
In this work we study in a general view slow rotating planets as Venus or
Titan which present superrotating winds in their atmospheres. We are interested
in understanding what mechanisms are candidates to be sources of net angular
momentum to generate this kind of dynamics.

In particular, in the case of Venus, in its atmosphere around an altitude of
100 Km relative to the surface, there exists winds that perform a full rotation around the planet in four terrestrial days, whereas the venusian day is
equivalent to 243 terrestrial ones. This phenomenon called superrotation is
known since many decades. However, its origin and behaviour is not completely
understood. In this article we analise and ponderate the importance of
different effects to generate this dynamics.

\end{abstract}
\volumeyear{year}
\volumenumber{number}
\issuenumber{number}
\eid{identifier}
\date[Date text]{date}
\received[Received text]{date}

\revised[Revised text]{date}

\accepted[Accepted text]{date}

\published[Published text]{date}

\startpage{101}
\endpage{102}
\maketitle
\tableofcontents

\section{Introduction}

Superrotation in Venus is an effect known since forty or more years. However,
the origin and mechanism of sustenance of these fast winds is still an open
question. In this report we want to review different proposals for the causes
of this phenomenon, and to introduce a simple model to clarify the basic
physical effects necessary to sustain the winds by sun irradiation. 

The upper atmosphere of a planet is regarded as the region of the atmosphere
whose structure and dynamics (temperature and density distribution,
composition and winds) are governed by the direct absorption of solar
radiation. The diversity and complexity of the processes taking place in the
upper atmospheres have made it necessary to develop their study into
a-separate branch of geophysics and astrophysics---aeronomy, which uses many
branches of physics and some branches of chemistry. The study of upper
atmospheres has both practical and theoretical interest. The knowledge of the
characteristics of the charged components of the upper atmosphere (which
constitutes the ionosphere of the planet) is needed to improve radio
communications and radio navigation (including space navigation); the
knowledge of the characteristics of the neutral upper atmosphere is needed to
determine the trajectories and lifetimes of artificial satellites and the
trajectories of space probes that enter the atmosphere. Clarification of the
mechanisms by which the influence of solar activity is transmitted through the
upper atmosphere to the troposphere is one of the important tasks in the
problem of solar-terrestrial relations. Comparative study of the upper
atmospheres of different planets and, in particular, of the dissipation of
gases from the atmospheres assists in clarifying the problem of the evolution
of planetary atmospheres. The dynamics of the Venus atmosphere presents a
major unsolved problem in planetary science: the so-called superrotation of
the lower atmosphere and its transition to solar-antisolar circulation in the
upper atmosphere. In general the dividing line between the lower and upper
atmosphere at 90--100 km altitude (pressure 0.39 to 0.028 mbar), the base of
the day-side thermosphere.) Superrotation has also been observed in the
atmosphere of Titan, the only other slowly rotating world with a substantial
atmosphere known at present. In this case also the transition to a different
circulation in the upper atmosphere is also apparent but not well understood.
Thus, the issues discussed below may be generic to any slowly rotating
terrestrial planet's atmosphere.

\subsection{Comparison with Earth}

Venus is one of the terrestrial planets, together with Earth and Mars, and has
similar mass and radius to those of Earth. Probably at the beginning of its
history it had a similar atmosphere to our planet. For unknown reasons Venus
has a slow retrograde rotation respect Earth. \begin{figure}[th]
\includegraphics[scale=0.45]{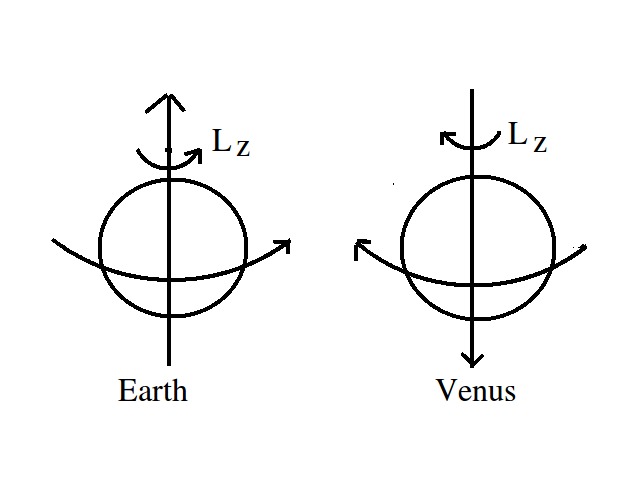}\caption{Sketch of retrograde
movement of Venus.}%
\end{figure}Apparently the magnetic dynamo in Venus is turned off, and thus
its atmosphere is completely exposed to the action of the solar wind, which
lead to several authors to speculate an initial loss of hydrogen and oxygen,
that in principle explains the absence of water. However, enough
concentrations of nitrogen and carbon exist to generates a greenhouse effect,
maintaining a dense and acid atmosphere compared with that of our planet.
Temperature varies along altitude, and due to superrotation there are not
strong fluctuations between day and night sides in the upper atmospheric
levels. One point to note is that the difference in Albedo between day and
night is extreme. The radiation in those regions is approximately that of a
black body and a white body, respectively. \begin{table}[thth]
\caption{Typical values of Earth and Venus \citep{2014ApJS..213...18P}.}%
\begin{tabular}
[c]{lccc}\hline\hline
Scale & Symbol & Venus & Earth\\\hline\hline
Radius (m) & $r$ & $6,051\cdot10^{6}$ & $6\cdot10^{6}$\\
Magnetic Field (G) & B & $\sim10^{-5}$ & 0,5\\
Rotation period (earth days) & T & 243 & 1\\
Equatioral surface velocity (m/s) & $v_{eq}$ & 1.8 & 465\\
Super Wind Periodicity (earth days) & $T_{sw}$ & 4 & -\\
Temperature Range (K) & T & \{228 - 773\} & \{184 -331\}\\
CO$_{2}$ (\%) &  & 96 & 0,04\\
N$_{2}$ (\%) &  & 3 & 78,1\\
Pressure at Troposphere level (Earth Atmosphere) & P & 92 & 1\\
Mass (Kg) & M & $4,9\cdot10^{24}$ & $5,98\cdot10^{24}$\\
Density ($g/cm^{3}$) & $\rho$ & 5,24 & 5,51\\
Days/Year (local) &  & 1,92 & 365,25\\
eccentricity &  & 0 & 0,016\\
Albedo &  & 0,65 & 38\\
Distance to the Sun (AU) &  & 0,723 & 1\\\hline
\end{tabular}
\caption{Sketch of retrograde movement of Venus.}%
\end{table}

\section{Early attempts}

In this section we describe some essential and basic characteristics of
Venusian at\-mos\-phere and its interaction with the surrounding ambient. The
atmosphere presents two dynamic regimes, zonal superrotation in the
troposphere and mesosphere \citep{1997veii.conf..459G} and solar-antisolar
circulation trough the terminator line in thermosphere
\citep{1997veii.conf..259B}, \citep{peralta}.

\subsection{Solar Wind Interaction}

\begin{figure}[th]
\includegraphics[scale=0.4]{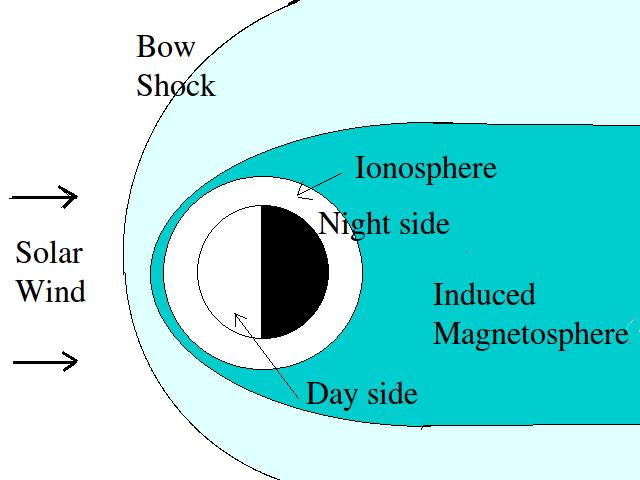}\caption{Scheme of the induced
magnetosphere structure of Venus due to the action of solar wind.}%
\label{mhd-venus}%
\end{figure}

Although Venus has apparently no magnetic field of its own, the solar wind in
the higher atmosphere levels generates an induced magnetosphere. The schematic
cartoon is shown in Figure \ref{mhd-venus}. 

\subsection{Atmospheric dynamics}

We call \textquotedblleft neutral atmosphere\textquotedblright\ that
corresponding to intermediate altitudes where the hydrodynamic approximation
is valid, \textit{i.e.} we can use the Navier-Stokes equations. Coriolis force
due to the slow rotation of Venus is negligible. It means that the
cyclostrophic approximation where pressure gradient is comparable to
centrifugal force, $\nabla p\sim f_{centrifuge}$, is valid, in opposition to
the Earth where the geostrophic approximation, in which the pressure gradient
is proportional to Coriolis force $\nabla p\sim f_{coriolis}$, is more appropriate.

The atmosphere is stratified with a layer of clouds around 100 Km of altitude
that has essentially different absorptions rates and stronger winds.
\begin{figure}[th]
\includegraphics[scale=0.35]{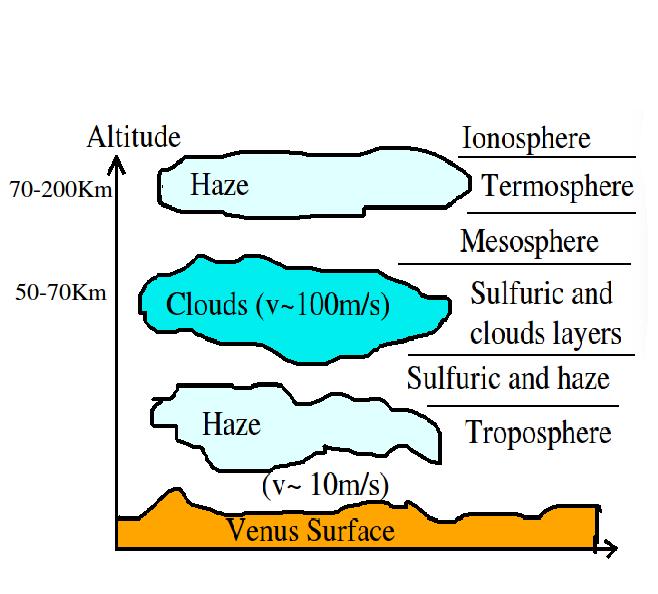}\caption{Cartoon of venusian
atmospherical levels. The presence of superrotating winds are at the cloud
layer.}%
\end{figure}In the clouds layer, speed winds are maximal, of the order of
120m/s, as well as the absorption of solar irradiance. At lower layers the
speed of the winds decrease abruptly and are close to co-rotation with the
planetary surface, sixty times below that in the cloud layer. Above the clouds
the wind velocity also decreases.

\begin{figure}[th]
\includegraphics[scale=0.6]{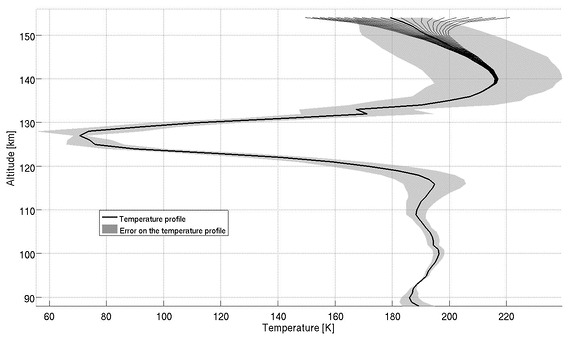}\caption{Temperature vs altitude
profile at Venus terminator. Continuous line is the temperature and grey
region is the error. \citep{2012JGRE..117.7001M}}%
\label{temp-profile}%
\end{figure}

Using the instrument SOIR on board the ESA Venus Express, Mahieux et al,
(2012) \citep{2012JGRE..117.7001M} measured different carbon dioxide densities
(the main component of the atmosphere) in the Venus terminator different
profiles. They established temperature profiles in this region for different
latitude and altitudes, some of the results are shown in Figure
\ref{temp-profile}. 

\subsection{Atmospheric Cells}

\begin{figure}[th]
\includegraphics[scale=0.45]{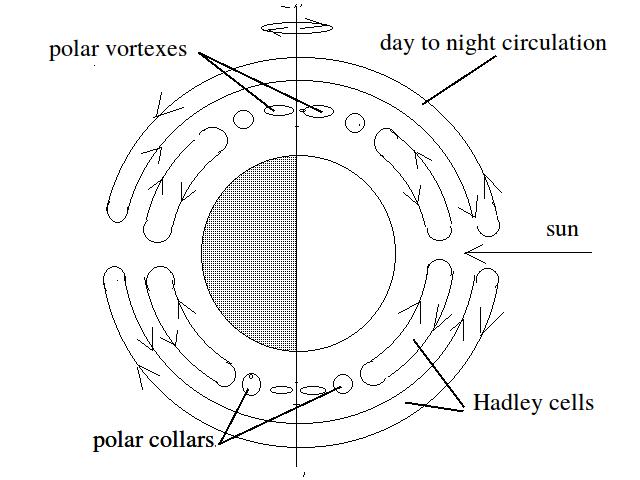}\caption{Scheme of Hadley
convective cells and superrotating winds.}%
\label{hadley}%
\end{figure}

These cells transports heat from the equator to the poles by convection, the
scheme is shown in Figure \ref{hadley}. The warm gas travels to the poles and
closes the cell transporting cold gas which toward the equator al low
altitudes. Essentially, there are four cells, two in the northern hemisphere
and two in the southern one. The two cells in each hemisphere have a
separatrix in the solar (midday) and anti-solar lines (midnight). Over these
cells, in the thermosphere, the circulation cells from solar to anti-solar
points takes place, the so called transterminator flow. Close to the poles in
the mesosphere polar vortices and polar collars are present.

\section{Superrotation}

To have superrotation, it is necessary to build up net angular momentum over
the whole atmosphere around the planet. Also, a stationary balance between
sources of angular momentum and diffusive effects to maintain an equilibrium
in time is necessary too. In this section we are going to mention some of
these effects to evaluate their role in this dynamics.

\subsection{Solar wind influence}

On Mars and Venus the solar wind (the flux of solar plasma) has a very
important influence on the atmosphere because of the weak magnetic fields of
these planets. According to the measurements, the magnetic fields of Mars and
Venus are about $10^{-4}$ of the terrestrial magnetic field; Mars apparently
has an intrinsic magnetic moment while the magnetic field of Venus is induced
by the effect of the solar wind on the ionosphere. \citep{dolginov1},
\citep{1969KosIs...7..747D}, \citep{dolginov2}, \citep{1967Sci...158.1669B},
\citep{1973SPhD...17.1117D} and \citep{1973JGR....78.4779D}.

The nature of the influence of the solar wind has not yet been completely
clarified, though the main features are as follows
\citep{1970AnRFM...2..313S}, \citep{1970P&SS...18.1281S},
\citep{1973P&SS...21..463C}, \citep{1974P&SS...22..967C} and
\citep{1973JGR....78.3169B}. The plasma of the solar wind (with concentrations
of a few particles per 1 cm$^{3}$, frozen magnetic field $\overline{B}%
_{sw}5-20nT$, and velocities of about 300-600 km/sec near Venus and Mars)
compresses the ionospheric plasma on the dayside, forming a sharp boundary---
the plasmopause or ionopause---while on the nightside it forms the tail of the
plasmosphere. The streamlines of the solar wind are deflected, so that it
flows round the plasmosphere. Above and along the flux a shock wave is formed,
in which the velocity, magnetic field, and temperature of the solar wind
change abruptly. The plasmopause on Venus was discovered at an altitude of
about 500 km by means of the radar-occultation experiments with Mariner 5. The
presence of a shock wave near Mars and Venus has been confirmed by magnetic
and plasma measurements by Venera 4 and 6, Mariner 4 and 5, and Mars 2 and 3
\citep{1965Sci...149.1241S}, \citep{Smith1}, \citep{gringauz1},
\citep{1970KosIs...8..431G}, \citep{1973Icar...18...54G}, \citep{gringauz2},
\citep{gringauz3} and \citep{1974KosIs..12..279V}. The shape of the shock wave
agrees satisfactorily with the one calculated in the framework of the
hypersonic gas-dynamic model.

The position of the plasmopause is determined in the hydrodynamic
approximation by the condition of balance of the dynamic pressure of the solar
wind and the pressure of the ionosphere:%
\begin{equation}
\varkappa n_{sw}m_{sw}v_{sw}\cos^{2}\left(  \zeta\right)  +\frac{\overline
{B}_{sw}^{2}}{8\pi}=k\left(  n_{e}T_{e}+n_{i}T_{i}\right)  +\frac{\overline
{B}^{2}}{8\pi}\label{f}%
\end{equation}
where $n_{sw}$, $m_{sw}$, $v_{sw}$, $B_{sw}$ are the concentration, the mean
mass of the particles, the velocity, and the magnetic field in the solar wind,
$\varkappa\sim$0.88 for the flow considered here; $n_{e}$, $n_{i}$, $T_{e}$,
$T_{i}$ are the concentrations and temperatures of the ionospheric electrons
and ions, and $\overline{B}$ is the magnetic field in the ionosphere; $\zeta$
is the angle between the outer normal to the mesopause and the velocity of the
unperturbed solar wind.

Note that neglecting the interaction between the solar wind and the neutral
particles simplifies the real picture since although the cross sections of
interaction of solar wind particles with the neutral particles are smaller
than those with the charged particles, there are many more neutral than
charged particles near the base of the exosphere.

From Eq. \ref{f} and using the experimental data on the height of the
plasmopause on Venus and parameters of the ionosphere and the solar wind it
was found that the magnetic field on Venus at altitude 500 km is $\overline
{B}\sim$ 20-30 nT, Cloutier and Daniell (1973) \cite{1973P&SS...21..463C}
specifying a model of the ionosphere and calculating the altitude distribution
of the conductivity and the currents induced by the solar wind and their
magnetic field, found the magnetopause as the altitude at which the induced
magnetic field becomes equal to the magnetic field of the solar wind. They
obtained a plasmopause at about 500 km on Venus and 350-425 km on Mars
(without allowance for its intrinsic magnetic moment). On the basis of this
model, it was then calculated \cite{1974P&SS...22..967C} that the influence of
the particles of the solar wind on the ionospheric particles above the
plasmopause leads to the latter being "swept out" of the atmosphere, and
although these losses are small (8 g/sec on Mars and 12 g/sec on Venus), the
profiles of the ionospheric ions and electrons above the plasmopause are
strongly distorted from the barometric distributions. Bauer and Hartlet (1973)
\cite{1973JGR....78.3169B} pointed out that Mars has an intrinsic magnetic
moment $\mu$= 2.4.\ 10$^{22}$ G/cm$^{3}$ (magnetic field on the surface
$\overline{B}_{0}$ = 60 nT), found by approximate estimates that the
magnetopause in the subsolar point is at the altitude 990 km (where
$\overline{B}\sim$ 20 nT) and that at about 300 km there is a plasmopause,
below which the ionospheric plasma is in hydrostatic equilibrium and rotates
with the planet, and above which there are large-scale convective cu\-rrents
of thermalized plasma induced by the solar wind. The possibility that
particles of the solar wind penetrate into the plasmosphere has not yet been
sufficiently studied. It has been shown in Whitten and Collin (1974)
\cite{1974RvGSP..12..155W} that the hydrodynamic relation (6), which
essentially determines the plasmopause as a wall that is impenetrable for
particles of the solar wind, is approximate; in reality, turbulence of the
plasma behind the shock wave may cause instabilities to arise on the
plasmopause, and these can allow particles of the solar wind to enter the
plasmosphere. On the basis of these ideas, an energy source was introduced in
some ionospheric models at the upper boundary of the ionosphere
\cite{1971P&SS...19..443H}. On the other hand, in Cloutier and Daniell (1974)
\cite{1974P&SS...22..967C}, also on the basis of a qualitative argument, it
was found that if particles of the solar wind pass through the plasmopause
electric forces arise which prevent their penetration into the plasmosphere
and return them to the outer flux. In the model of \cite{1974P&SS...22..967C}
there is a sink on the upper boundary of the ionosphere due to the ionospheric
particles being "swept out" by the solar wind.

As was pointed out in Izakov (2001) \cite{iz} a further study of the
interaction of the solar wind with the atmosphere is needed in order to make
more precise the upper boundary conditions in theoretical models of the
atmosphere and the ionosphere.

\subsection{Gierasch mechanism}

One of the most firm candidates to explain the superrotation of Venus
atmosphere is Gierasch mechanism (GM), (Gierasch, 1975 \cite{gierasch75}).
In GM angular momentum is generated in the lower atmosphere, for instance, by
the drag of the planet surface, and a Hadley cell convects it upward near the
equator. The meridional flow of the cell at high altitudes transport angular
momentum toward the poles, near of which it is convected downward. GM also
requires the existence of a mechanism that opposes the poleward advection at
high altitudes, that is, an enhanced horizontal diffusion of angular momentum
in the upper levels. If there is also limited vertical diffusion, a net
accumulation of angular momentum takes place at high altitude in
mid-latitudes, thus leading to superrotation. The conditions for GM to work
are rather restrictive: a large Richardson number is necessary in order to
suppress vertical transport by instabilities. This in turn requires a thermal
balance between radiative heating and adiabatic cooling due to vertical
transport. This radiative-convective equilibrium leads to a net flux of heat
from equator to poles to compensate for the resulting lack of radiative
equilibrium. Since heat is transported poleward by the cell flow and opposed
by horizontal heat diffusion, a large enough meridional flow is required,
while at the same time horizontal diffusion of angular momentum must dominate
over the transport by the flow.

It is thus not surprising that early Global Circulation Models \ (GCM) applied
to Venus failed to generate superrotation, apparently due to large vertical
diffusion by thermally unstable atmosphere profiles (Del Genio A. D. and
Souzzo R. J., 1987, \cite{delgenio87}).
GCM that incorporate more realistic atmosphere profiles do indeed show
superrotation of the right cha\-rac\-te\-ris\-tics (Yamamoto M. and Takahashi
M., 2003, \cite{yamamoto2003}),
and favor the GM for its ge\-ne\-ra\-tion, with the development of a single,
large Hadley cell, and enhanced horizontal diffusion due mainly to fast
gravity waves and slower Rossby waves.

\subsection{Other candidates to generate superrotating winds}

The superrotation is allocated in the termosphere, the transterminator flow is
in the mesosphere, and the Hadley cells are allocated in the troposphere. The
transterminator flow could be an agent to propagate momentum between different
layers of the atmosphere, adding thermal convection and the rotation of the
planet could be a promising candidate to generates superrotating winds. \citep{2010arXiv1005.3488D}

\subsubsection{Simple Model for an external forcing}

We can compute the torque due an external forcing, for example, solar wind,
reducing to 2D problem, considering radial incidence:
\begin{equation}
\vec{\tau}=\vec{r}\times\vec{f(t)}=r\hat{r}\times f(t)(\cos\alpha\hat{r}%
-\sin\alpha\hat{\alpha})=-rf(t)\sin\alpha\hat{z}%
\end{equation}
\begin{figure}[th]
\includegraphics[scale=0.34]{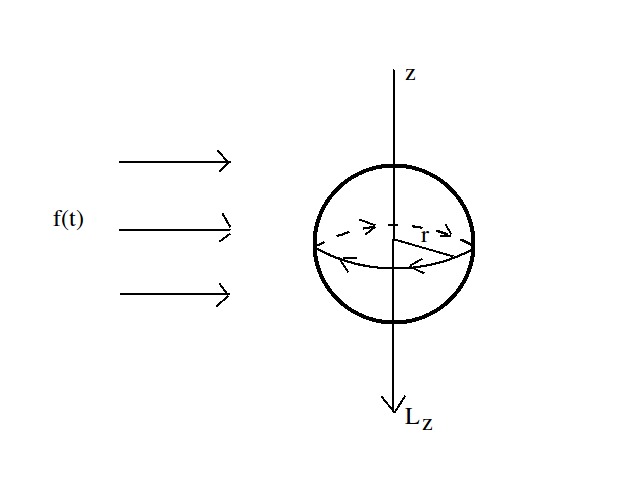}
\includegraphics[scale=0.34]{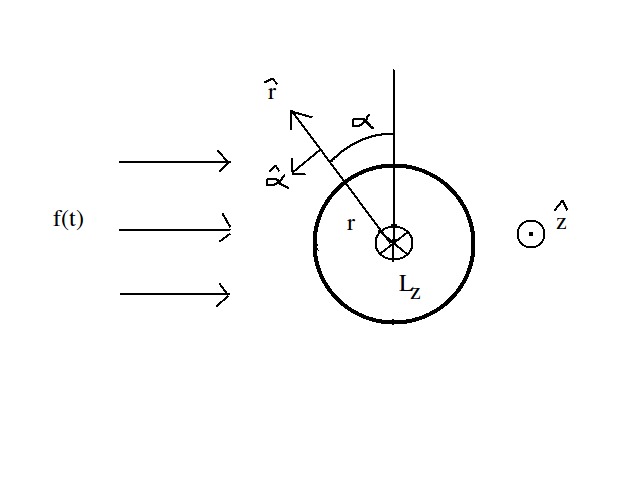}\caption{\textbf{Left:} Cartoon to
schematic torque computation generated by an external forcing. \textbf{Right:}
Azimuthal projection.}%
\end{figure}Using the fact that $\displaystyle\tau=\frac{dL}{dt}$, we have:
\begin{equation}
\ell_{z}=-\int r~f(t)\sin\alpha~dt
\end{equation}
Looking at a given $r$ and $\alpha$ we obtain:
\begin{equation}
\ell_{z}=-r~\sin\alpha~\int f(t)dt
\end{equation}
By definition $p=mv\Rightarrow\delta p=fdt$ and $\Delta p=\int\delta p$.
Therefore:
\begin{equation}
\ell_{z}=-r~\sin\alpha~\Delta p
\end{equation}
If we integrate over every $\alpha$, by symmetry the only contribution will be
for $\alpha=\pi/2$
\begin{equation}
\ell_{z}=-r~\Delta p
\end{equation}
It means that for an external contribution to the atmospherics winds, we need
a non-zero net moment transfer.\newline\newline But, is it enough? In
principle, this seems to be difficult because by symmetry the only point that
receive this contribution is the solar point which is a \textquotedblleft
point\textquotedblright\ of zero measure. In any case, if there is an external
source of superrotation wind, it is necessary to show temporal and velocity
scales, dispersion relation and waves propagation due this effect.

\subsubsection{Temperature Gradients}

As mentioned before, between day and night sides there exist strong gradients
of temperature. In principle in the upper levels the gradient is smoother than
in superficial levels, because superrotating winds tends to uniformise the
temperature differences more efficiently than in the lower levels, where the
winds are slower by an order of magnitude. \citep{2010arXiv1005.3488D}
\begin{figure}[th]
\includegraphics[scale=0.3]{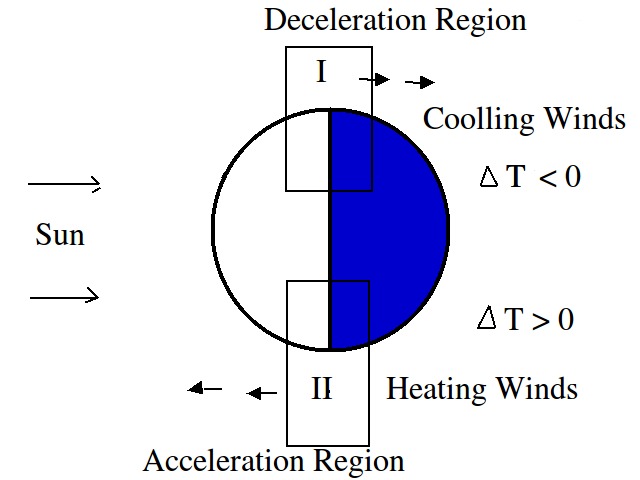}
\includegraphics[scale=0.3]{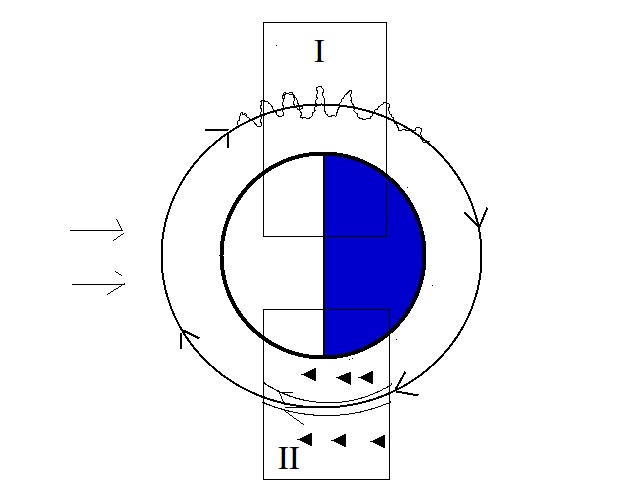}\caption{\textbf{Left:} Temperature
gradients provocates two regions, Region I; where the winds are stopped, the
morphology is corrugated clouds shape. In opposite, Region II; the winds are
accelerated, corresponding to smooth clouds shape. \textbf{Right:} Cartoon of
wind shape in each region.}%
\label{shape}%
\end{figure}

\subsubsection{Albedo Radiation}

\begin{figure}[th]
\includegraphics[scale=0.35]{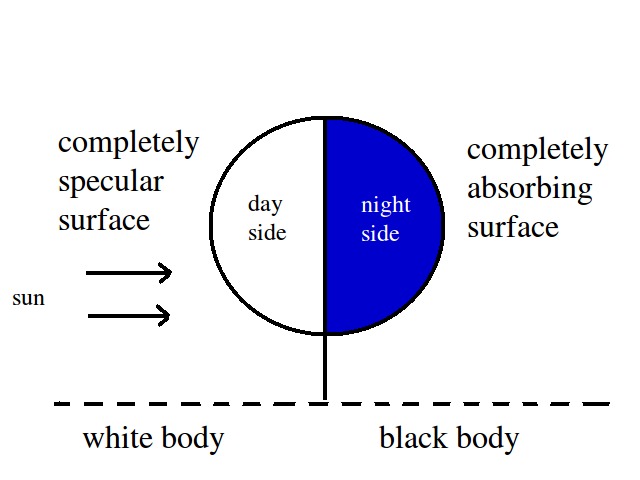}\caption{The Albedo difference
between day side and night side generates a pressure gradient from day to
night.}%
\label{blackbody}%
\end{figure}

Venus is a slow rotating planet, and the day and night sides have a strong
difference in their Albedo. The behaviour of the incident radiation on the
planet can be approximated as a white body in the day side and like a black
body in the nightside, see Figure \ref{blackbody}. Chafin (2014)
\citep{2014arXiv1406.0116C} estimated this pressure gradient like
\begin{equation}
\tau=\int z~dF=\int z~P_{rad}~dA=\int_{0}^{r}z~P_{rad}~(2\sqrt{R^{2}-z^{2}%
})~dz=\frac{2}{3}P_{rad}~r^{3}%
\end{equation}
with $P_{rad}\propto9\cdot10^{-6}$Pa. He showed that this value is compatible
with superrotating wind.~\vspace*{-0.75cm}\newline

\subsubsection{Durand Manterola's hypothesis}

Between altitudes in between 150 and 800km supersonic winds exist called
transterminator flows. Considering again Fig \ref{hadley}, the external cell
due to the influence of solar wind generates a photo-ionization pressure
gradient. There two cells from day to night sides. However, the planet also
slowly rotates. As shown in Fig \ref{shape} two regions can be defined:

\begin{itemize}
\item \textit{Region I: dusk zone;} the external layer of the cell has a
solidary speed respect to the rotation of the planet with velocity $v_{I}$.

\item \textit{Region II: dawn zone;} the external layer of the cell has an
opposite speed respect to the rotation of the planet with velocity $v_{II}$.
\end{itemize}

\begin{figure}[th]
\includegraphics[scale=0.3]{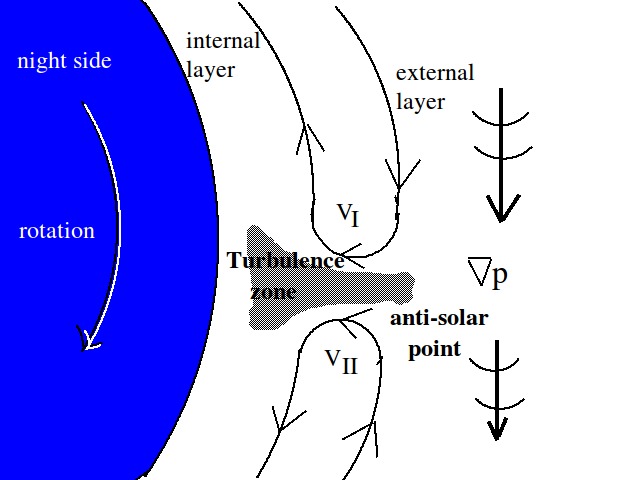}\caption{Presence of transterminator
flows, in the anti-solar point could generate a transversal gradient of
pressure which colaborates with the angular momentum of superrotating winds. }%
\label{durand}%
\end{figure}

\begin{figure}[thth]
\includegraphics[scale=0.33]{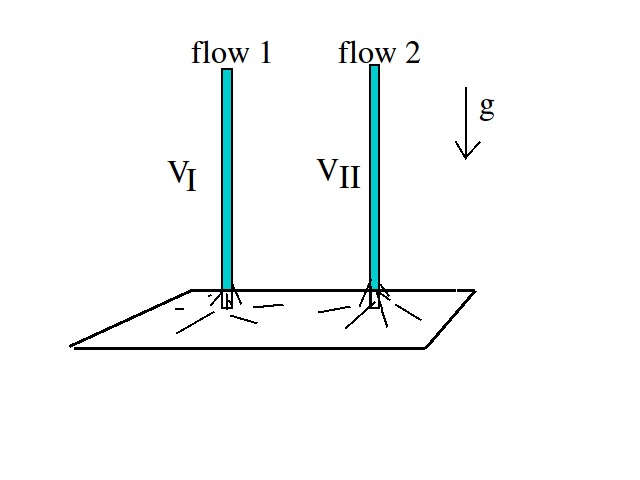}
\includegraphics[scale=0.33]{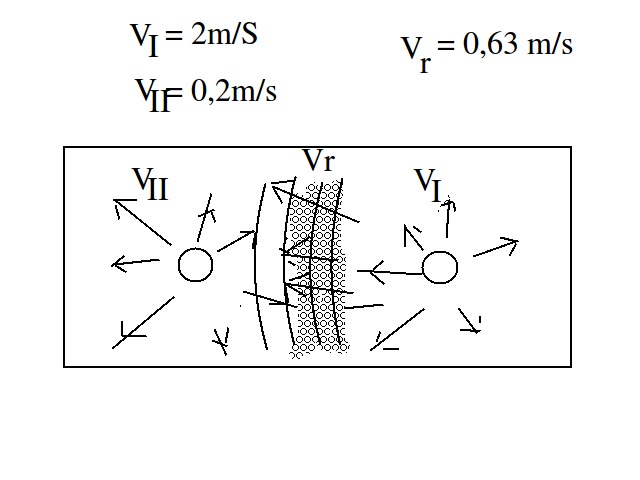}\caption{Cartoon of
Durand-Manterola's experiment which try to emulate the Fig \ref{durand}.
\textbf{Left:} in perspective. \textbf{Right:} azimuthal view. }%
\label{experiment}%
\end{figure}

In the night side, particularly in the anti-solar point, one can consider the
mechanism in Fig \ref{durand}. In this region $v_{I}=10V_{II}$ the contact of
the two flows generates turbulence and a pressure gradient which creates waves
transferring angular momentum in the retrograde sense, between external layers
of Hadley cells and superrotating winds in the upper levels. This energy,
according to \cite{2010arXiv1005.3488D} has enough power to overcome viscosity
losses and could explain superrotating winds. \cite{2010arXiv1005.3488D} found that using the Darcy-Weisbach
equation, two fluxes at different velocities generate turbulence and wave
propagation, transferring energy between different atmospheric layers:
\begin{equation}
\delta P=f_{D}~\frac{L}{D}~\frac{1}{2}~\rho v^{2}%
\end{equation}
where $f_{D}$ is the friction, $D$ the pipe diameter, $L$ typical
longitude.\newline They set an experiment with two water \textquotedblleft
pipes\textquotedblright\ falling off at different velocities (2m/s and 0.2m/s)
and found wave propagation and a the generation of a net flow (velocity
0,63m/s) giving a net angular momentum. The basic scheme of the experiment is
shown in Fig \ref{experiment} 

\section{Gravity waves}

As mentioned above in the Gierasch mechanism the role of gravity waves is
relevant to the generation of superrotation. For that reason, it is important
to observe and measure their behaviour. 

Gravity waves are wave disturbances in which buoyancy acts as the restoring
force. These waves (internal gravity or buoyancy waves) abound in the stable
density layering of the upper atmosphere. Their effects are visibly manifest in
the curls of the clouds, in the moving skein-like and billow patterns of the
clouds in the middle of the venusian atmosphere, and in the slowly shifting
bands of the top part of the atmosphere. 
In terrestrial planets the characteristic of gravity waves are similar and its
description is analogous to gravity waves in Earth, and since the seventies
detailed studies exist about it, for example see Francis (1975)
\cite{1975JATP...37.1011F}.

What produces them? These waves can be generated by disturbances in the lowest part of the atmosphere, for example, wind flow over mountain ranges and violent thunderstorms. Jet stream shear and solar radiation are other sources. An initial small amplitude in the lower atmosphere increases with height until the waves break in the mesosphere and lower thermosphere. Their wavelengths can range up to some hundreds of kilometers. Their periods ranging from a few minutes to days.
Given the possible generation by flow over mountain ranges  \cite{2014Icar..227...94P}, detailed works exist relating wave morphology and mapping of the planets, for example see Basilevsky (2003) \cite{2003RPPh...66.1699B}.

Recent works, Fukuhara et al. (2017) \cite{fukuhara}, using the orbiter
instrument Akatsuki suggest that bow shaped structures are the result of
gravity waves generated in the lower atmosphere by mountain topography (around
10 km height), that propagate upwards. The authors modeled large scale gravity waves to be compared with observations supporting these assumptions. 
Although the dayside is well known, the nightside is not sufficiently observed. Peralta et al. (2017)
\cite{peralta-nat} using results from the Venus Express mission report that stationary waves in the upper atmosphere in the nightside are slower than in the dayside hemisphere, imposing constraints to Venus general circulation models that do not predict such phenomena. Akatsuki and Venus Express observations will be important to elucidate the features in the upper clouds in the day and night sides.

\section{Simple Model}

\begin{figure}[ptb]
\includegraphics[scale=0.3]{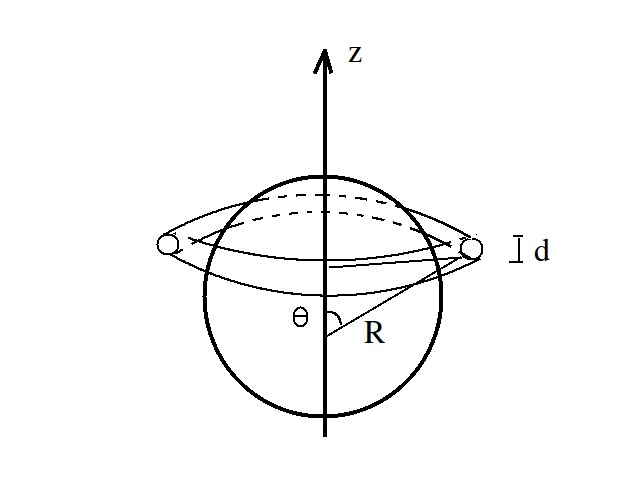}
\includegraphics[scale=0.3]{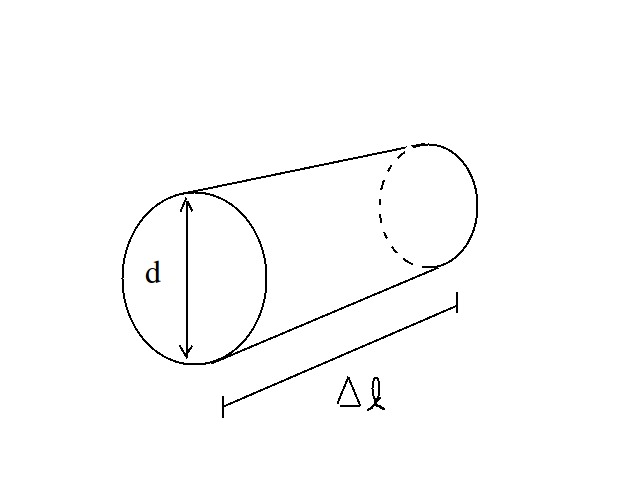}\caption{Cylindrical volume of
diameter $d=\lambda_{abs}$, at radius $R$ and angle $\theta$ with dimensions
$\delta Sup=\pi\lambda_{abs}\Delta\ell\Rightarrow\delta V=\frac{\pi
\lambda_{abs}^{2}}{4}\Delta\ell$}%
\label{dV}%
\end{figure}

In this section we are going to show that superrotation speed solutions are
possible in a stationary regime due a temperature gradient by the sun
irradiation in a slowly rotating planet. If we consider a differential volume
at given latitude and altitude, see Figure \ref{dV}.

\subsection{Basic equations}

To explore the possibility to generate superrotating winds by the temperature
gradient due to the solar irradiation over the atmosphere, we build a
simplified model which has the following assumptions.

\begin{itemize}
\item Stationary wind regime.

\item $2\pi$-periodicity condition in the solar point.

\item Friction, radial and azimuthal velocities are negligible compared to
$u_{\phi}$, the zonal wind velocity.

\item We take the atmosphere as an ideal gas.
\end{itemize}

The continuity equation is then:
\begin{equation}
\frac{\partial}{\partial\phi}(\rho u_{\phi})=0
\end{equation}
Where $\rho$ is the mass density.\newline\newline The dynamic equations are:
\begin{equation}
\frac{u_{\phi}}{r}\frac{\partial u_{\phi}}{\partial\phi}=-\frac{1}{\rho}%
\frac{\partial p}{\partial\phi}%
\end{equation}
where $r\sin\theta=R$, and
\begin{equation}
\frac{u_{\phi}^{2}}{2}=-\frac{1}{\rho}\frac{\partial p}{\partial r}-g_{r}%
\end{equation}
where $g_{r}$ is the radial gravity component. Also, the specific entropy $s$
equation is:
\begin{equation}
\rho\frac{u_{\phi}}{r}\frac{\partial s}{\partial\phi}=\frac{\dot{Q}}{T}%
\end{equation}
Where $\dot{Q}$ is the rate of absorbed or emitted heat per unit volume, and
$T$ the temperature. Under the assumption of ideal gas and using the first
principle of the thermodynamics:
\begin{equation}
\delta Q-\delta W=dU\Rightarrow T\delta S-p\delta V=nC_{v}dT
\end{equation}
where $n$ is the number of mol and $R_{m}=0.082\ell\mathrm{{atm}/molK}$ we
obtain:
\begin{equation}
\delta s=C_{v}~\frac{dT}{T}-R_{m}\frac{\delta\rho}{\rho}\label{entropy}%
\end{equation}
Combining the energy equation with the entropy equation:
\begin{equation}
\frac{R_{m}T}{\rho}\frac{\partial\rho}{\partial\phi}=C_{v}\frac{\partial
T}{\partial\phi}-\frac{r\dot{Q}}{\rho u_{\phi}}\label{balance}%
\end{equation}
Since $\delta p=R_{m}\rho dT+R_{m}T\delta\rho$ we have the equation for radial
velocity:
\begin{equation}
\frac{1}{2}\frac{\partial u_{\phi}^{2}}{\partial\phi}=u_{\phi}\frac{\partial
u_{\phi}}{\partial\phi}=-R_{m}\frac{\partial T}{\partial\phi}-R_{m}\frac
{T}{\rho}\frac{\partial\rho}{\partial\phi}%
\end{equation}
Using the continuity equation, $R_{m}=C_{p}-C_{v}$ and constant $\rho u_{\phi
}$, we have a balance equation like:
\begin{equation}
{\frac{\partial}{\partial\phi}(C_{p}T+u_{\phi}^{2}/2)=\frac{r\dot{Q}}{\rho
u_{\phi}}}\label{qdot}%
\end{equation}
Also, $\dot{Q}$ is the balance between the albedo radiation, black-body
emission and diffusivity:
\begin{equation}
\dot{Q}=\dot{Q}_{\odot}-\sigma T^{4}\frac{\delta Sup}{\delta V}+\frac{k}%
{r^{2}}\frac{\partial^{2}T}{\partial\phi^{2}}%
\end{equation}
Where
\begin{equation}
\dot{Q}_{\odot}=\left\{
\begin{array}
[c]{ll}%
\displaystyle\frac{I_{\odot}}{4}\frac{\delta Sup}{\delta V}~f_{abs} &
\mathrm{{if}~~~\phi>\phi_{0}}\\
0 & \mathrm{{if}~~~\phi\leq\phi_{0}}%
\end{array}
\right.  \nonumber
\end{equation}
Considering an annular volume and neglecting the diffusion term, we have:
\begin{equation}
{\dot{Q}=\frac{I_{\odot}}{\lambda_{abs}}sen\theta cos\phi~f_{abs}%
\Theta(cos\phi)-\frac{4\sigma}{\lambda_{abs}}(T^{4}-T_{ref}^{4})}\nonumber
\end{equation}
being:

\begin{itemize}
\item $f_{abs}$ an absorption factor related with Albedo which depends on the
temperature and solar radiation.

\item $\Theta(cos\phi)$ Heaviside function.

\item $I_{\odot}$ solar radiative intensity.

\item $\lambda_{abs}$ characteristic absorption scale.

\item $T_{ref}$ is a reference temperature.
\end{itemize}

Using again the continuity equation in the expression for velocity we arrive
at the following relation:
\begin{equation}
u_{\phi}+\frac{p}{\rho u_{\phi}}=cte
\end{equation}
Finally, assuming ideal gas...
\begin{equation}
\frac{dT}{d\phi}=\frac{r\dot{Q}}{c_{p}\rho u_{\phi}}=\frac{r\dot{Q}}{c_{p}%
K}\sqrt{\frac{KR_{m}T}{p_{0}}},
\end{equation}
where
\begin{equation}
K=\sqrt{\frac{pu_{\phi}^{2}}{R_{m}T}}%
\end{equation}
is the eigenvalue obtained by imposing a $2\pi$ periodic solution.

\subsection{Numerical experiments}

With the model developed in last section we took observed values of pressure
and temperature as a function of altitude to evaluate wind speeds and compare
with the observed ones.

According to our model, preserving the periodicity and fitting parameters
adequately, we obtained wind velocity profiles in the range between 50 and 80
Km similar to those observed. In this way we obtained each profile for a given
latitude and altitude for the whole planet. We made this for Venus and Titan.

\subsubsection{Venus}

\begin{figure}[ptb]
\includegraphics[scale=0.45]{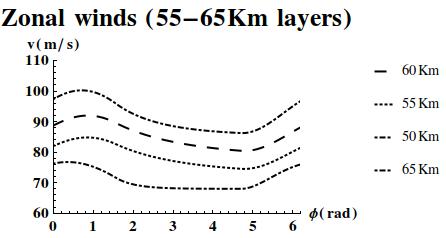}\qquad
\includegraphics[scale=0.45]{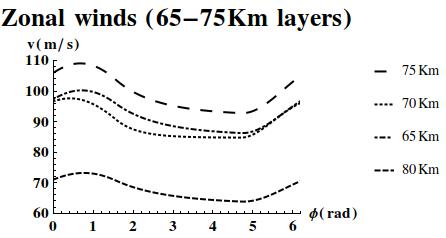}\caption{Zonal winds at different 
altitudes versus latitude, $\phi=0$ midday, $\phi=\pi$ midnight. Left: Between 
55 and 65 Km layers. Right: Between 65 and 75 Km layers. 
}
\label{venus-wind}%
\end{figure}From Figure \ref{venus-wind} we can see that the speed is lower in
the interval $\{\pi/3;5\pi/3\}$ co\-rres\-pon\-ding to the migration from the
solar point to the nightside $\{\pi/2;3\pi/2\}$. Later, when the wind returns
to the dayside the speed increases. \begin{figure}[ptbptb]
\includegraphics[scale=0.45]{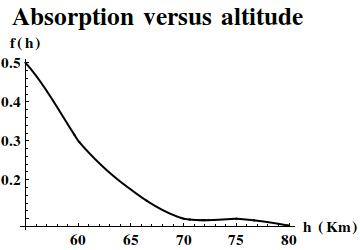}\qquad
\includegraphics[scale=0.45]{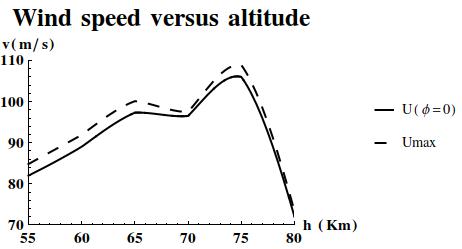}\caption{Left: absorption
  coefficients used for numerical fitting. Right: Comparison between speed wind
  at $\phi=0$ and maximum speed for the layers in the range 50-80km height}%
\label{abs}%
\end{figure}Fig \ref{abs} shows the absorption coefficients as a function of
altitude used for numerical fitting (left) and the speed of wind in the solar
point comparing with the maximal speed (right). We can see for one side that it
which is consistent with the fact that the clouds layer increases the
absorption and by another side the difference in velocity in each layers are not
greater than a $10\%$.

\subsubsection{Titan}

In the same way we made the fit taking atmospheric values of Titan, in a
qualitative way, we observed a similar behaviour to the Venusian case. Comparing
with Venus the difference in the fluctuations in speed are lower due essentially
by the solar irradiance. Qualitatively, the values are similar to the observed
ones indicating that the influence of Saturn it seems not relevant to the wind
dynamics for our model.
\begin{figure}[ptb]
\includegraphics[scale=0.45]{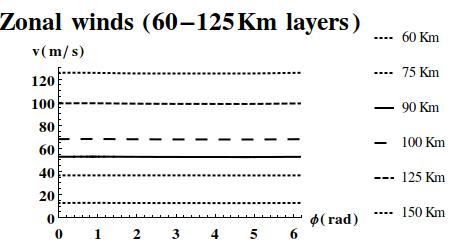}\qquad
\includegraphics[scale=0.43]{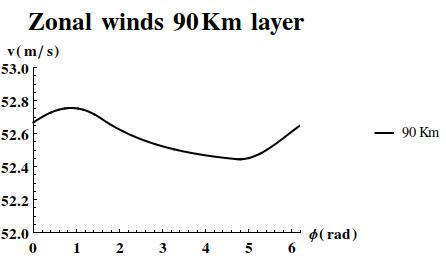}\caption{Zonal winds at different 
altitudes versus latitude, $\phi=0$ midday, $\phi=\pi$ midnight. Left: Between 
60 and 125 Km layers. Right: More detailed zonal wind at 90Km layer.}%
\end{figure} \begin{figure}[ptbptb]
  ~\hspace{-6cm}
  \includegraphics[height=3.9cm, width=5.5cm]{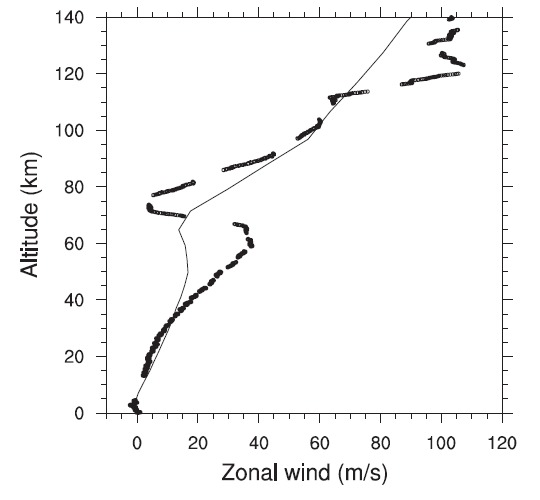}\vspace{-3.9cm}\\
 ~\hspace{6cm} 
\includegraphics[scale=0.21]{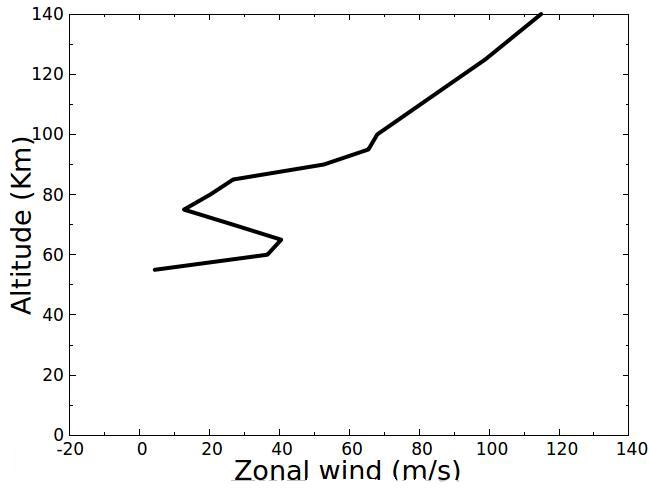}\caption{Left: observed values
  (Bird et al \cite{2005Natur.438..800B}). Right: values obtained with our
  model.}%
\end{figure}

\section{Concluding remarks an outlook}

In this paper we have reviewed and analyzed the problem of superrotation on
Venus in light of the interest that the latest data have produced in the
scientific community. These data introduce a clear indication that traditional
theoretical approaches have to be modified in some way. Although the model
introduced here is capable of reproducing the general behavior of the wind in
super rotation (at least in the layers of interest), under the assumptions of
slow rotating planet and stationary regime due by heat balance is stablished.
The model indicates that the main source to supply superrotation is the solar
irradiance. Ho\-we\-ver, the problem of the generation of superrotation remains
an enigma. Another problem is described in \cite{fukuhara} that is interesting
from the theoretical point of view because it presents the challenge of
constructing a formulation that gives a global description of the dynamics of
the Venusian atmosphere. The other question is whether a relationship between
the phenomenon of superrotation on different planets and satellites certainly
exists. 

\section{Acknowledgements}
We gratefully acknowledge to the Departamento de F\'\i sica (FCEyN, UBA) and
Instituto de F\'\i sica del Plasma (CONICET-UBA), FM and DC are also grateful
to the Consejo Nacional de Investigaciones Cient\'\i ficas y T\'ecnicas, CV is
also grateful to the Instituto de Ciencias (UNGS) and DC is also grateful to the
Bogoliubov Laboratory of Theoretical Physics for their institutional support.
\bibliography{biblio}
\end{document}